# Rortex – A New Vortex Vector Definition and Vorticity Tensor and Vector Decompositions

Chaoqun Liu[1,a)], Yisheng Gao[1], Shuling Tian[2,1] Xiangrui Dong[3,1]

[1]*Department of Mathematics, University of Texas at Arlington, Arlington, Texas 76019, USA*

[2]*College of Aerospace Engineering, Nanjing University of Aeronautics and Astronautics, Nanjing, Jiangsu, 210016, China*

[3]*National Key Laboratory of Transient Physics, Nanjing University of Science & Technology, Nanjing, Jiangsu, 210094, China*

A vortex is intuitively recognized as the rotational/swirling motion of the fluids. However, an unambiguous and universally-accepted definition for vortex is yet to be achieved in the field of fluid mechanics, which is probably one of the major obstacles causing considerable confusions and misunderstandings in turbulence research. In our previous work, a new vector quantity which is called vortex vector was proposed to accurately describe the local fluid rotation and clearly display vortical structures. In this paper, the definition of the vortex vector, named Rortex here, is revisited from the mathematical perspective. The existence of the rotational axis is proved through real Schur decomposition. Based on real Schur decomposition, a fast algorithm for calculating Rortex is also presented. In addition, new vorticity tensor and vector decompositions are introduced: the vorticity tensor is decomposed to a rigidly rotational part and an anti-symmetric deformation part, and the vorticity vector is decomposed to a rigidly rotational vector and a non-rotational vector. Several cases, including 2D Couette flow, 2D rigid rotational flow and 3D boundary layer transition on a flat plate, are studied to demonstrate the justification of the definition of Rortex. It can be observed that Rortex identifies both the precise swirling strength and the rotational axis,

[a)]Email: cliu@uta.edu

and thus it can reasonably represent the local fluid rotation and provide a new powerful tool for vortex dynamics and turbulence research.

## Nomenclature

| | | |
|---|---|---|
| $g_Z$ | = | the criterion of rotation of fluid element |
| $g_{Z\theta}$ | = | the criterion of rotation of fluid element at an azimuth $\theta$ |
| $\boldsymbol{P}$ | = | transformation matrix for rotating around Z axis |
| $Q$ | = | second invariant of velocity gradient tensor, a criterion of vortex identification |
| $\boldsymbol{Q}$ | = | transformation matrix |
| $\vec{r}$ | = | vector of direction of rotation in frame $xyz$ |
| $\vec{R}$ | = | Rortex (vortex vector) |
| $\vec{S}$ | = | non-rotational part of vorticity vector |
| $u, v, w$ | = | components of the velocity vector in frame $xyz$ |
| $U, V, W$ | = | components of the velocity vector in frame $XYZ$ |
| $U', V'$ | = | components of velocity in $X'Y'$ plane |
| $\boldsymbol{v}$ | = | velocity vector in frame $xyz$ |
| $\boldsymbol{V}$ | = | velocity vector in frame $XYZ$ |
| $x, y, z$ | = | coordinates in the reference frame $xyz$ |
| $X, Y, Z$ | = | coordinates in the reference frame $XYZ$ |
| $X', Y'$ | = | coordinates in the reference frame $X'Y'$ obtained by rotating the $XY$ around Z |
| $\alpha$ | = | local strain rate of fluid element |
| $\beta$ | = | averaged angular velocity of fluid element |
| $\varphi$ | = | the angle of phase |
| $\lambda_{ci}$ | = | a criterion of vortex identification, imaginary part of the complex conjugate eigenvalue of velocity gradient tensor |
| $\lambda_2$ | = | a criterion of vortex identification, second-largest eigenvalue. |
| $\xi_Z$ | = | component of vorticity in Z |
| Ω | = | a criterion of vortex identification, the ratio of vorticity over the whole motion of fluid element |
| $\nabla \vec{v}$ | = | velocity gradient tensor in the frame *xyz* |
| $\nabla \vec{V}$ | = | velocity gradient tensor in the frame *XYZ* |
| $\nabla \vec{V}_\theta$ | = | velocity gradient tensor in the frame rotated from *XYZ* by an angle θ |
| $\nabla \times \vec{v}$ | = | the curl of velocity $\vec{v}$, vorticity |
| $|\cdot|$ | = | the magnitude of a vector |

## I. INTRODUCTION

Vortices are ubiquitous in nature and technology, may be observed from whirlpools to tornadoes, from bilge vortices underneath a ship's hull to wingtip vortices behind a wing.[1] Some vortical structures, such as hairpin vortices, referred as coherent turbulent structures,[2] are recognized as one of the most important characteristics of turbulent flow and have been studied for more than 60 years.[3] It is generally acknowledged that intuitively, vortices represent the rotational/swirling motion of the fluids. However, a precise and rational definition of vortex is deceptively complicated and remains an open issue.[4,5]

In classical vortex dynamics,[4,6,7] vortex is usually associated with vorticity which has a rigorous mathematical definition (curl of velocity). Wu et al. define vortex as "a connected region with high concentration of vorticity compared with its surrounding".[4] Lamb uses vorticity tubes to define vortices.[8] Nitsche asserts that "A vortex is commonly associated with the rotational motion of fluid around a common centerline. It is defined by the vorticity in the fluid, which measures the rate of local fluid rotation."[9] While these vorticity-based vortex definitions are straightforward, they are not always satisfying. An immediate contradiction to these definitions is that the Blasius boundary layer where vorticity is large near the wall, but no rotation/swirl motion (considered as vortex) is observed, as vorticity cannot distinguish a vortical region from a shear layer region. Another common problem in practice is the determination of an appropriate threshold above which vorticity can be considered as high concentrated to identify vortices. The threshold is subjective and arbitrary.[10] In addition, the maximum vorticity does not necessarily represent the center of the vortex. Robinson has pointed out that the association between regions of strong vorticity and actual vortices can be rather weak in the turbulent boundary layer, especially in the near wall region.[11] Wang et al. also find that vorticity magnitude will be reduced when vorticity lines enter the vortex region and vorticity magnitude inside the vortex region is much smaller than the surrounding area, especially near the solid wall in a flat plate boundary layer, for most three-dimensional vortices like Λ-shaped vortices.[12]

Another possible candidate for vortex definition is the one based on closed or spiraling streamlines.[13] Robinson et al. claim that "A vortex exists when instantaneous streamlines mapped onto a plane normal to the vortex core exhibit a roughly circular or spiral pattern, when viewed from a reference frame moving with the center of the vortex core."[14] Although it seems intuitive, Lugt has pointed out that "the definition and identification of a vortex in unsteady motions is difficult since streamlines and pathlines are not invariant with respect to Galilean and rotational transformations. Recirculatory streamline patterns at a certain instant in time do not necessarily represent vortex motions in which fluid particles are moving around a common axis. Thus, instantaneous streamline patterns do not provide enough information to be used for the definition of a vortex."[13]

Due to the essential requirement for visualizing vortical structures and their evolution in turbulence, several vortex identification criteria have been developed, including $\lambda_2$-criterion,[15] Q-criterion,[16] $\lambda_{ci}$-criterion[17] and $\lambda_{cr}/\lambda_{ci}$-criterion,[18] etc. Nevertheless, these methods still fail to provide a quantitative definition of vortices. Moreover, these methods require proper thresholds. It is difficult to determine which threshold is proper, since different thresholds will indicate different vortical structures. For example, even if the same DNS data on the late boundary layer transition is employed, vortex breakdown will be exposed for some large threshold in Q-criterion while no vortex breakdown can be found for some smaller threshold. This will directly influence one's understanding and explanation on the mechanism of turbulence generation, i.e. turbulence is caused by vortex breakdown or not caused by vortex breakdown. Recently, a new vortex identification method called Ω-method is proposed by Liu et al., based on the idea that vortex is a connected area where vorticity overtakes deformation.[19] Ω-method possesses several advantages, such as no need for a case-related threshold, easy to perform and able to capture strong and weak vortices simultaneously. But it is still not the ideal answer to the question of the definition of vortex, owing to some limitations such as the introduction of an artificial parameter ε and the incapability to identify the swirl axis and its orientation. Kolář formulates a triple decomposition from which the residual vorticity can be obtained after the extraction of an effective pure shearing motion and represents a direct measure of the pure rigid-body rotation of a fluid element.[20] However, the triple decomposition is not unique, so a so-called basic reference frame must be first determined. Searching for the basic reference frame results in an expensive optimization problem for every point in the flow field, which limits the applicability of the method. Kolář et al. also introduce the concepts of the maximum corotation and the average corotation of line segments near a point and apply these methods for vortex identification.[21,22] However, the so-called maximum-corotation method suffers from the unstable behavior of the maxima and the averaged corotation vector is evaluated by integration over a unit sphere, which makes it difficult to be used to study the transport property of vortex. In addition to the mentioned Eulerian vortex identification methods, some objective Lagrangian vortex identification methods are developed to study the vortex structures in the rotating reference frame.[23,24] For more details on currently available vortex identification methods, one can refer to recent review papers by Zhang et al.[25] and Epps.[10]

As the rigorous definition of vortex is a key issue to understand vortex and turbulence[29,30], a definition of vortex vector was proposed to describe the local fluid rotation in our previous work.[26] In this paper, the vortex vector, which is newly named Rortex to avoid confusion with common terminologies, is revisited from the mathematical perspective. A rigorous mathematical proof of the existence of the rotational axis is presented by real Schur decomposition. Upon real Schur decomposition, a fast algorithm for calculating Rortex, which dramatically improves the computational efficiency when compared to our previous method, is developed. Moreover, the introduction of Rortex allows us performing new vorticity tensor and vorticity vector decompositions: the vorticity tensor is decomposed to a rigidly rotational part and an anti-symmetric deformation part, and the vorticity vector is decomposed to a rigidly rotational vector ($\vec{R}$) and a non-rotational vector ($\vec{S}$), i.e. $\nabla \times \vec{V} = \vec{R} + \vec{S}$. It should be noted that although Kolář has developed the similar ideas in a series of papers[20-22] and the definition of Rortex is equivalent to Kolář's residual vorticity in 2D cases, the direction of the swirling axis and the swirling strength in our definition are totally different from Kolář's results in 3D cases since we first identify the direction of the swirl axis and then determine the swirling strength while Kolář uses extremum seeking or average process to find both magnitude of swirling and direction of the swirl axis. It is unclear if the direction obtained by Kolář's extremum seeking or average process represents the actual direction of rotation, since neither vector nor vector line plots are provided in Kolář's papers and one cannot visualize the direction of the swirl axis.

The paper is organized as follows. The definition of Rortex is revisited in Section II. The proof of the existence of the rotational axis and the fast algorithm for calculating Rortex are presented in Section III. In Section IV, new vorticity tensor decomposition is given. Based on the tensor decomposition, vorticity vector is decomposed into Rortex vector and non-rotational shear vector in Section Ⅴ. Section Ⅵ shows several 2D and 3D examples to justify our definition of Rortex. The conclusion remarks are summarized in the last section.

## II. REVISIT OF THE DEFINITION OF RORTEX (VORTEX VECTOR)

In this section, our previous work on the definition of the vortex vector Rortex is briefly presented.[26] Some new mathematical proofs are provided here, instead of the original descriptions.

### A. Determination of the direction of the rotational axis

In our previous work, we transform the original $xyz$-frame to a new reference $XYZ$-frame in which the fluid-rotational axis vector $\vec{r}$ is parallel to the $Z$-axis. The velocity gradient tensor $\nabla \vec{v}$ and $\nabla \vec{V}$ in reference frame $xyz$ and $XYZ$ can be written respectively as follows,

$$\nabla \vec{v} = \begin{bmatrix} \frac{\partial u}{\partial x} & \frac{\partial u}{\partial y} & \frac{\partial u}{\partial z} \\ \frac{\partial v}{\partial x} & \frac{\partial v}{\partial y} & \frac{\partial v}{\partial z} \\ \frac{\partial w}{\partial x} & \frac{\partial w}{\partial y} & \frac{\partial w}{\partial z} \end{bmatrix}, \qquad \nabla \vec{V} = \begin{bmatrix} \frac{\partial U}{\partial X} & \frac{\partial U}{\partial Y} & \frac{\partial U}{\partial Z} \\ \frac{\partial V}{\partial X} & \frac{\partial V}{\partial Y} & \frac{\partial V}{\partial Z} \\ \frac{\partial W}{\partial X} & \frac{\partial W}{\partial Y} & \frac{\partial W}{\partial Z} \end{bmatrix}$$

There is a relation between these two coordinate systems:

$$\nabla \vec{V} = Q \nabla \vec{v} Q^{-1} \tag{1}$$

where $Q$ is the transformation matrix from the $xyz$-frame to the $XYZ$-frame. The necessary condition for a velocity gradient tensor to have rotation in the $Z$-axis only is

$$\frac{\partial U}{\partial Z} = 0 \tag{2}$$

$$\frac{\partial V}{\partial Z} = 0 \tag{3}$$

This leads to a new transformed the $XYZ$ frame where

$$\nabla \vec{V} = \begin{bmatrix} \frac{\partial U}{\partial X} & \frac{\partial U}{\partial Y} & 0 \\ \frac{\partial V}{\partial X} & \frac{\partial V}{\partial Y} & 0 \\ \frac{\partial W}{\partial X} & \frac{\partial W}{\partial Y} & \frac{\partial W}{\partial Z} \end{bmatrix} \tag{4}$$

In our previous work, we have addressed that the criteria of fluid rotation around axis $R_i$ is $\frac{\partial U_j}{\partial X_k} \cdot \frac{\partial U_k}{\partial X_j} > 0, i = 1, 2, 3$. There is no rotation around $R_i$ if $\frac{\partial U_j}{\partial X_k} \cdot \frac{\partial U_k}{\partial X_j} \leq 0, i = 1, 2, 3$.

**Theorem 1**: If a tensor has $\frac{\partial U}{\partial Z} = 0$, there is no rotation around the $Y$ axis.

Proof:

$$\begin{bmatrix} \frac{\partial U}{\partial X} & \frac{\partial U}{\partial Y} & 0 \\ \frac{\partial V}{\partial X} & \frac{\partial V}{\partial Y} & \frac{\partial V}{\partial Z} \\ \frac{\partial W}{\partial X} & \frac{\partial W}{\partial Y} & \frac{\partial W}{\partial Z} \end{bmatrix} = \begin{bmatrix} \frac{\partial U}{\partial X} & \frac{1}{2}\left(\frac{\partial U}{\partial Y} + \frac{\partial V}{\partial X}\right) & \frac{1}{2}\frac{\partial W}{\partial X} \\ \frac{1}{2}\left(\frac{\partial V}{\partial X} + \frac{\partial U}{\partial Y}\right) & \frac{\partial V}{\partial Y} & \frac{1}{2}\left(\frac{\partial W}{\partial Y} + \frac{\partial V}{\partial Z}\right) \\ \frac{1}{2}\frac{\partial W}{\partial X} & \frac{1}{2}\left(\frac{\partial W}{\partial Y} + \frac{\partial V}{\partial Z}\right) & \frac{\partial W}{\partial Z} \end{bmatrix} + \begin{bmatrix} 0 & -\frac{1}{2}\left(\frac{\partial V}{\partial X} - \frac{\partial U}{\partial Y}\right) & -\frac{1}{2}\frac{\partial W}{\partial X} \\ \frac{1}{2}\left(\frac{\partial V}{\partial X} - \frac{\partial U}{\partial Y}\right) & 0 & -\frac{1}{2}\left(\frac{\partial W}{\partial Y} - \frac{\partial V}{\partial Z}\right) \\ \frac{1}{2}\frac{\partial W}{\partial X} & \frac{1}{2}\left(\frac{\partial W}{\partial Y} - \frac{\partial V}{\partial Z}\right) & 0 \end{bmatrix} \tag{5}$$

$$B = \begin{bmatrix} 0 & -\frac{1}{2}\left(\frac{\partial V}{\partial X} - \frac{\partial U}{\partial Y}\right) & -\frac{1}{2}\frac{\partial W}{\partial X} \\ \frac{1}{2}\left(\frac{\partial V}{\partial X} - \frac{\partial U}{\partial Y}\right) & 0 & -\frac{1}{2}\left(\frac{\partial W}{\partial Y} - \frac{\partial V}{\partial Z}\right) \\ \frac{1}{2}\frac{\partial W}{\partial X} & \frac{1}{2}\left(\frac{\partial W}{\partial Y} - \frac{\partial V}{\partial Z}\right) & 0 \end{bmatrix} = \begin{bmatrix} 0 & 0 & -\frac{1}{2}\frac{\partial W}{\partial X} \\ 0 & 0 & 0 \\ \frac{1}{2}\frac{\partial W}{\partial X} & 0 & 0 \end{bmatrix} + \begin{bmatrix} 0 & -\frac{1}{2}\left(\frac{\partial V}{\partial X} - \frac{\partial U}{\partial Y}\right) & 0 \\ \frac{1}{2}\left(\frac{\partial V}{\partial X} - \frac{\partial U}{\partial Y}\right) & 0 & 0 \\ 0 & 0 & 0 \end{bmatrix} +$$

$$\begin{bmatrix} 0 & 0 & 0 \\ 0 & 0 & -\frac{1}{2}\left(\frac{\partial W}{\partial Y} - \frac{\partial V}{\partial Z}\right) \\ 0 & \frac{1}{2}\left(\frac{\partial W}{\partial Y} - \frac{\partial V}{\partial Z}\right) & 0 \end{bmatrix} = C + D + E \tag{6}$$

The rotation criteria is $R_i = -\frac{\partial U_j}{\partial X_k} \cdot \frac{\partial U_k}{\partial X_j} > 0$ and then $R_Y = \frac{\partial U}{\partial Z} \cdot \frac{\partial W}{\partial X} = 0 \cdot \frac{\partial W}{\partial X} = 0$. According to the rotation criteria, in tensor $C$, $R_Y = 0$, $R_X = 0$, $R_Z = 0$, and there is no rotation in any coordinate axis like axes $X, Y$, or $Z$. In tensor $\boldsymbol{D}$, $R_Y = 0$, $R_X = 0$ and in tensor $\boldsymbol{E}$, $R_Y = 0$, $R_Z = 0$. Therefore, for tensor $\boldsymbol{B}$, there is no rotation around the $Y$-axis.

Similarly, if a tensor has $\frac{\partial V}{\partial Z} = 0$, there is no rotation around the $X$-axis.

**Corollary 1**: If a tensor has $\frac{\partial U}{\partial Z} = 0$ and $\frac{\partial V}{\partial Z} = 0$, there is no rotation around the $Y$-axis and the $X$-axis. There is a possible rotation around the $Z$-axis which is the only possible rotational axis.

Proof:

$$\begin{bmatrix} \frac{\partial U}{\partial X} & \frac{\partial U}{\partial Y} & 0 \\ \frac{\partial V}{\partial X} & \frac{\partial V}{\partial Y} & 0 \\ \frac{\partial W}{\partial X} & \frac{\partial W}{\partial Y} & \frac{\partial W}{\partial Z} \end{bmatrix} = \begin{bmatrix} \frac{\partial U}{\partial X} & \frac{1}{2}\left(\frac{\partial V}{\partial X}+\frac{\partial U}{\partial Y}\right) & \frac{1}{2}\frac{\partial W}{\partial X} \\ \frac{1}{2}\left(\frac{\partial V}{\partial X}+\frac{\partial U}{\partial Y}\right) & \frac{\partial V}{\partial Y} & \frac{1}{2}\frac{\partial W}{\partial Y} \\ \frac{1}{2}\frac{\partial W}{\partial X} & \frac{1}{2}\frac{\partial W}{\partial Y} & \frac{\partial W}{\partial Z} \end{bmatrix} + \begin{bmatrix} 0 & -\frac{1}{2}\left(\frac{\partial V}{\partial X}-\frac{\partial U}{\partial Y}\right) & -\frac{1}{2}\frac{\partial W}{\partial X} \\ \frac{1}{2}\left(\frac{\partial V}{\partial X}-\frac{\partial U}{\partial Y}\right) & 0 & -\frac{1}{2}\frac{\partial W}{\partial Y} \\ \frac{1}{2}\frac{\partial W}{\partial X} & \frac{1}{2}\frac{\partial W}{\partial Y} & 0 \end{bmatrix} = \boldsymbol{A} + \boldsymbol{B} \quad (7)$$

$$\boldsymbol{B} = \begin{bmatrix} 0 & -\frac{1}{2}\left(\frac{\partial V}{\partial X}-\frac{\partial U}{\partial Y}\right) & -\frac{1}{2}\frac{\partial W}{\partial X} \\ \frac{1}{2}\left(\frac{\partial V}{\partial X}-\frac{\partial U}{\partial Y}\right) & 0 & -\frac{1}{2}\frac{\partial W}{\partial Y} \\ \frac{1}{2}\frac{\partial W}{\partial X} & \frac{1}{2}\frac{\partial W}{\partial Y} & 0 \end{bmatrix} = \begin{bmatrix} 0 & -\frac{1}{2}\left(\frac{\partial V}{\partial X}-\frac{\partial U}{\partial Y}\right) & 0 \\ \frac{1}{2}\left(\frac{\partial V}{\partial X}-\frac{\partial U}{\partial Y}\right) & 0 & 0 \\ 0 & 0 & 0 \end{bmatrix} + \begin{bmatrix} 0 & 0 & -\frac{1}{2}\frac{\partial W}{\partial X} \\ 0 & 0 & 0 \\ \frac{1}{2}\frac{\partial W}{\partial X} & 0 & 0 \end{bmatrix} +$$

$$\begin{bmatrix} 0 & 0 & 0 \\ 0 & 0 & -\frac{1}{2}\frac{\partial W}{\partial Y} \\ 0 & \frac{1}{2}\frac{\partial W}{\partial Y} & 0 \end{bmatrix} = \boldsymbol{C} + \boldsymbol{D} + \boldsymbol{E} \quad (8)$$

There is only tensor $\boldsymbol{C}$ which possibly has rotation while tensors $\boldsymbol{D}$ and $\boldsymbol{E}$ have no rotation.

After $\boldsymbol{Q}$ tensor transformation, we can find that we only have a possible rotational vector $\boldsymbol{r} = Z$.

### B. Rotation strength in the 2D plane

Through $\boldsymbol{Q}$ tensor transformation, we have found the possible rotational axis which is $\boldsymbol{r} = Z$, but we still need to find the rotation strength which should be $2\min\left\{\left|\frac{\partial U}{\partial Y}\right|, \left|\frac{\partial V}{\partial X}\right|\right\}$ which represents the rigid rotation without any deformation. This can be achieved by a tensor transformation in plane $XY$, which we called P tensor transformation. As known, vorticity $\xi_Z = \frac{\partial V}{\partial X} - \frac{\partial U}{\partial Y}$ in a 2D flow is a Galilean invariant quantity, but $\frac{\partial V}{\partial X}$ or $\frac{\partial U}{\partial Y}$ is not invariant and will change with the rotation of the reference frame. When frame $XYZ$ is rotated around the $Z$-axis by an angle $\theta$, the new velocity gradient tensor is

$$\nabla\vec{V}_\theta = \boldsymbol{P}\nabla\vec{V}\boldsymbol{P}^{-1} \quad (9)$$

where $P$ is the transformation matrix around the Z axis and there are

$$P = \begin{bmatrix} \cos\theta & \sin\theta & 0 \\ -\sin\theta & \cos\theta & 0 \\ 0 & 0 & 1 \end{bmatrix}, \qquad P^{-1} = \begin{bmatrix} \cos\theta & -\sin\theta & 0 \\ \sin\theta & \cos\theta & 0 \\ 0 & 0 & 1 \end{bmatrix}$$

So, we can obtain

$$\frac{\partial V}{\partial X}|_\theta = \alpha \sin(2\theta + \varphi) + \beta \tag{10}$$

$$\frac{\partial U}{\partial Y}|_\theta = \alpha \sin(2\theta + \varphi) - \beta \tag{11}$$

where

$$\alpha = \frac{1}{2}\sqrt{\left(\frac{\partial V}{\partial Y} - \frac{\partial U}{\partial X}\right)^2 + \left(\frac{\partial V}{\partial X} + \frac{\partial U}{\partial Y}\right)^2} \tag{12}$$

$$\beta = \frac{1}{2}\left(\frac{\partial V}{\partial X} - \frac{\partial U}{\partial Y}\right) \tag{13}$$

and

$$\tan\varphi = \frac{\frac{\partial V}{\partial X} + \frac{\partial U}{\partial Y}}{\frac{\partial V}{\partial Y} - \frac{\partial U}{\partial X}} \tag{14}$$

Actually $\beta = \xi_Z = \frac{\partial V}{\partial X} - \frac{\partial U}{\partial Y}$ is a vorticity component.

According to our criteria for fluid rotation, we must have

$$g_{Z\theta} = -\frac{\partial V}{\partial X}|_\theta \frac{\partial U}{\partial Y}|_\theta = \beta^2 - \alpha^2 \sin^2(2\theta + \varphi) > 0 \tag{15}$$

In order to satisfy this condition for all $\theta$, we must require $\beta^2 > \alpha^2$. In order to decompose the tensor $C$ to be a rigid rotation, it must be required to get $\min\left\{\left|\frac{\partial U}{\partial Y}|_\theta\right|, \left|\frac{\partial V}{\partial X}|_\theta\right|\right\}$ for all $\theta$. Since $\alpha$ is always positive, if $\beta > 0$, $\min\left\{\left|\frac{\partial U}{\partial Y}|_\theta\right|, \left|\frac{\partial V}{\partial X}|_\theta\right|\right\} = \frac{\partial V}{\partial X}|_{\theta\min} = \beta - \alpha$ and the rigid rotation strength is $R = 2\frac{\partial V}{\partial X}|_{\theta\min} = 2(\beta - \alpha)$. For simplicity, we write $R = 2\frac{\partial V}{\partial X}$ which should be understood as $R = 2\frac{\partial V}{\partial X}|_{\theta\min}$ after the P rotation.

If $\beta < 0$, $\min\left\{\left|\frac{\partial U}{\partial Y}|_\theta\right|, \left|\frac{\partial V}{\partial X}|_\theta\right|\right\} = \frac{\partial U}{\partial Y}|_{\theta\min} = -(\beta + \alpha)$ and then $R = 2\frac{\partial U}{\partial Y}|_{\theta\min}$.

We now get both the direction $r$ and the rigid rotation strength and are then ready to give the definition for Rortex which is a rigid rotation part of the fluid motion at a local point P.

**C. Definition of Rortex and Vortex**

**Definition 1**: Rortex is defined as $\vec{R} = 2R\vec{r}$ which is a rigid rotation part of fluid motion at local point P.

**Definition 2**: Vortex is a connected area where $R \neq 0$.

## III. EXISTENCE PROOF AND FAST ALGORITHM BASED ON REAL SCHUR DECOMPOSITION

### A. Existence of the axis of rotation $\vec{r}$

We prove the existence theorem of the axis of rotation $\vec{r}$ using the real Schur decomposition. The real Schur decomposition theorem can be written as follows: [24]

If $A \in \mathbb{R}^{n \times n}$, then there exists an orthogonal $Q \in \mathbb{R}^{n \times n}$ such that

$$Q^{T}AQ = \begin{bmatrix} R_{11} & R_{12} & \cdots & R_{1m} \\ 0 & R_{22} & \cdots & R_{2m} \\ \vdots & \vdots & \ddots & \vdots \\ 0 & 0 & \cdots & R_{mm} \end{bmatrix} \tag{16}$$

Where each $R_{ii}$ is either a 1-by-1 real matrix or a 2-by-2 matrix which has complex conjugate eigenvalues.

**Theorem 2**. For any velocity gradient tensor $\nabla \vec{v}$, there exists a proper rotation matrix $Q$, such that

$$\nabla \vec{V} = Q \nabla \vec{v} Q^{-1} = \begin{bmatrix} \frac{\partial U}{\partial X} & \frac{\partial U}{\partial Y} & \frac{\partial U}{\partial Z} \\ \frac{\partial V}{\partial X} & \frac{\partial V}{\partial Y} & \frac{\partial V}{\partial Z} \\ \frac{\partial W}{\partial X} & \frac{\partial W}{\partial Y} & \frac{\partial W}{\partial Z} \end{bmatrix}, \frac{\partial U}{\partial Z} = 0, \frac{\partial V}{\partial Z} = 0 \tag{17}$$

$\frac{\partial U}{\partial X}, \frac{\partial U}{\partial Y}, \ldots$ are the components of the velocity gradient tensor $\nabla \vec{V}$ under the rotation. And the new direction of the Z-axis is the axis of rotation $\vec{r}$.

Proof. For any $3 \times 3$ real matrix $A$, it can be classified according to the eigenvalues: (a) $A$ has three real eigenvalues; (b) $A$ has one real eigenvalue and two complex conjugate eigenvalues.

Let

$$A = \nabla \vec{v}^T \tag{18}$$

If $A$ has three real eigenvalues, according to the real Schur decomposition, there exists an orthogonal matrix $Q^*$ and the Schur form $S$, such that

$$S = Q^{*T}AQ^* = \begin{bmatrix} R_{11} & R_{12} & R_{13} \\ 0 & R_{22} & R_{23} \\ 0 & 0 & R_{33} \end{bmatrix} \tag{19}$$

$R_{11}, R_{12}, \ldots, R_{33}$ represent $1 \times 1$ real matrices (real numbers) and $R_{11}, R_{22}, R_{33}$ are three real eigenvalues.

Since $Q^*$ is an orthogonal matrix, the determinant of $Q^*$ is either 1 or -1.

If the determinant of $Q^*$ is 1, let

$$Q = Q^{*T} \tag{20}$$

Then, $Q$ is also an orthogonal matrix (the transpose of an orthogonal matrix is orthogonal) and the determinant of $Q$ is 1 (the determinant of a matrix is the same as the determinant of its transpose). $Q$ is a proper rotation matrix. And we have

$$\nabla \vec{V} = Q\nabla\vec{v}Q^{-1} = QA^TQ^T = (QAQ^T)^T = (Q^{*T}AQ^*)^T = (S)^T = \begin{bmatrix} R_{11} & R_{12} & R_{13} \\ 0 & R_{22} & R_{23} \\ 0 & 0 & R_{33} \end{bmatrix}^T = \begin{bmatrix} R_{11} & 0 & 0 \\ R_{12} & R_{22} & 0 \\ R_{13} & R_{23} & R_{33} \end{bmatrix} \tag{21}$$

which meets the conditions of $\frac{\partial U}{\partial Z} = 0, \frac{\partial V}{\partial Z} = 0$.

If the determinant of $Q^*$ is -1, $Q^*$ represents an improper rotation (reflection). Let

$$Q = \begin{bmatrix} 1 & 0 & 0 \\ 0 & 1 & 0 \\ 0 & 0 & -1 \end{bmatrix} Q^{*T} \tag{22}$$

Then $Q$ is a proper rotation matrix. And we have

$$\nabla \vec{V} = Q\nabla\vec{v}Q^{-1} = QA^TQ^T = (QAQ^T)^T = \left(\begin{bmatrix} 1 & 0 & 0 \\ 0 & 1 & 0 \\ 0 & 0 & -1 \end{bmatrix} Q^{*T}AQ^* \begin{bmatrix} 1 & 0 & 0 \\ 0 & 1 & 0 \\ 0 & 0 & -1 \end{bmatrix}\right)^T =$$

$$\left(\begin{bmatrix} 1 & 0 & 0 \\ 0 & 1 & 0 \\ 0 & 0 & -1 \end{bmatrix} S \begin{bmatrix} 1 & 0 & 0 \\ 0 & 1 & 0 \\ 0 & 0 & -1 \end{bmatrix}\right)^T = \left(\begin{bmatrix} 1 & 0 & 0 \\ 0 & 1 & 0 \\ 0 & 0 & -1 \end{bmatrix} \begin{bmatrix} R_{11} & R_{12} & R_{13} \\ 0 & R_{22} & R_{23} \\ 0 & 0 & R_{33} \end{bmatrix} \begin{bmatrix} 1 & 0 & 0 \\ 0 & 1 & 0 \\ 0 & 0 & -1 \end{bmatrix}\right)^T =$$

$$\begin{bmatrix} R_{11} & R_{12} & -R_{13} \\ 0 & R_{22} & -R_{23} \\ 0 & 0 & R_{33} \end{bmatrix}^T = \begin{bmatrix} R_{11} & 0 & 0 \\ R_{12} & R_{22} & 0 \\ -R_{13} & -R_{23} & R_{33} \end{bmatrix} \tag{23}$$

which also meets the conditions of $\frac{\partial U}{\partial Z} = 0, \frac{\partial V}{\partial Z} = 0$.

If $A$ has one real eigenvalue and two complex conjugate eigenvalues, according to the real Schur decomposition, there exists an orthogonal matrix $Q^*$ and the Schur form $S$, such that

$$S = Q^{*T}AQ^* = \begin{bmatrix} R_{11} & R_{12} \\ 0 & R_{22} \end{bmatrix} = \begin{bmatrix} r_{11} & r_{12} & r_{13} \\ r_{21} & r_{22} & r_{23} \\ 0 & 0 & r_{33} \end{bmatrix} \tag{24}$$

where $R_{11} = \begin{bmatrix} r_{11} & r_{12} \\ r_{21} & r_{22} \end{bmatrix}$ is corresponding to two complex conjugate eigenvalues and $R_{22} = r_{33}$ is a real eigenvalue (we place the two complex conjugate eigenvalues on the top left of the Schur form). $R_{12} = \begin{bmatrix} r_{13} \\ r_{23} \end{bmatrix}$ represents the off-diagonal block.

If the determinant of $Q^*$ is 1, let

$$Q = Q^{*T}$$

Then, $Q$ is a proper rotation matrix. And we have

$$\nabla \vec{V} = Q\nabla\vec{v}Q^{-1} = QA^TQ^T = (QAQ^T)^T = (Q^{*T}AQ^*)^T = (S)^T = \begin{bmatrix} r_{11} & r_{12} & r_{13} \\ r_{21} & r_{22} & r_{23} \\ 0 & 0 & r_{33} \end{bmatrix}^T = \begin{bmatrix} r_{11} & r_{21} & 0 \\ r_{12} & r_{22} & 0 \\ r_{13} & r_{23} & r_{33} \end{bmatrix}$$

(25)

which meets the conditions of $\frac{\partial U}{\partial Z} = 0, \frac{\partial V}{\partial Z} = 0$.

If the determinant of $Q^*$ is -1, let

$$Q = \begin{bmatrix} 1 & 0 & 0 \\ 0 & 1 & 0 \\ 0 & 0 & -1 \end{bmatrix} Q^{*T}$$

Then $Q$ is a proper rotation matrix. And we have

$$\nabla \vec{V} = Q\nabla\vec{v}Q^{-1} = QA^TQ^T = (QAQ^T)^T = \left(\begin{bmatrix} 1 & 0 & 0 \\ 0 & 1 & 0 \\ 0 & 0 & -1 \end{bmatrix} Q^{*T}AQ^* \begin{bmatrix} 1 & 0 & 0 \\ 0 & 1 & 0 \\ 0 & 0 & -1 \end{bmatrix}\right)^T =$$

$$\left(\begin{bmatrix} 1 & 0 & 0 \\ 0 & 1 & 0 \\ 0 & 0 & -1 \end{bmatrix} \begin{bmatrix} r_{11} & r_{12} & r_{13} \\ r_{21} & r_{22} & r_{23} \\ 0 & 0 & r_{33} \end{bmatrix} \begin{bmatrix} 1 & 0 & 0 \\ 0 & 1 & 0 \\ 0 & 0 & -1 \end{bmatrix}\right)^T = \left(\begin{bmatrix} 1 & 0 & 0 \\ 0 & 1 & 0 \\ 0 & 0 & -1 \end{bmatrix} S \begin{bmatrix} 1 & 0 & 0 \\ 0 & 1 & 0 \\ 0 & 0 & -1 \end{bmatrix}\right)^T =$$

$$\begin{bmatrix} r_{11} & r_{12} & -r_{13} \\ r_{21} & r_{22} & -r_{23} \\ 0 & 0 & r_{33} \end{bmatrix}^T = \begin{bmatrix} r_{11} & r_{21} & 0 \\ r_{12} & r_{22} & 0 \\ -r_{13} & -r_{23} & r_{33} \end{bmatrix} \tag{26}$$

which also meets the conditions of $\frac{\partial U}{\partial Z} = 0, \frac{\partial V}{\partial Z} = 0$.

The axis of rotation $\vec{r}$ can be obtained using

$$\vec{r} = Q^T \begin{bmatrix} 0 \\ 0 \\ 1 \end{bmatrix} \quad (27)$$

Two notes should be given:

(a) Definition of the proper/improper rotation matrix. A proper rotation matrix is an orthogonal matrix with determinant 1. An improper rotation matrix is an orthogonal matrix with determinant -1. The proper rotation does not change the orientation of the coordinate system, so the right-handed Cartesian coordinate system is still right-handed under the proper rotation. But, the improper rotation changes the orientation of the coordinate system, e.g., from the right-handed Cartesian coordinate system to the left-handed Cartesian coordinate system.

In fact, the above proof is irrelevant to the orientation of the coordinate system, both proper and improper rotation are valid. Here we only use the proper rotation because usually we use right-handed Cartesian coordinate system and the proper rotation does not change the orientation.

(b) For any vector $\vec{x}$

$$\vec{X} = Q\vec{x}$$

represents the new vector under the rotation. The direction of the z-axis in the frame xyz can be written as

$$\vec{z} = \begin{bmatrix} 0 \\ 0 \\ 1 \end{bmatrix} \quad (28)$$

So the new direction of the z-axis under the rotation is

$$\vec{Z} = Q^T \begin{bmatrix} 0 \\ 0 \\ 1 \end{bmatrix} \quad (29)$$

which is the axis of rotation $\vec{r}$.

**B. Fast algorithm for calculating Rortex**

LAPACK provides a subroutine DGEES to perform real Schur decomposition. DGEES can output both the Schur form and the orthogonal matrix. It can also order eigenvalues by setting a select function so that selected eigenvalues are at the top left of the Schur form.

The complete procedure for calculating the vortex vector $\vec{R}$ is given in the following steps:

1) Compute the velocity gradient tensor $\nabla \vec{v}$ in the original frame xyz and $A = \nabla \vec{v}^T$;

2) Call the subroutine DGEES to obtain the orthogonal matrix $Q^*$ and the Schur form $S$, using a select function to sort the eigenvalues so that two complex conjugate eigenvalues are located at the top left of the Schur form;

3) Calculate the determinant of $Q^*$. If the determinant is 1, use eq. (6) to obtain the proper rotation matrix $Q$. If the determinant is -1, use eq. (8) to obtain the proper rotation matrix $Q$;

4) Calculate the axis of rotation $\vec{r}$ using eq. (10) and the velocity gradient tensor $\nabla \vec{V}$ in the frame XYZ using eq. (7), (9), (11), (12);

5) Calculate $\alpha$ and $\beta$ using

$$\alpha = \frac{1}{2}\sqrt{\left(\frac{\partial V}{\partial Y} - \frac{\partial U}{\partial X}\right)^2 + \left(\frac{\partial V}{\partial X} + \frac{\partial U}{\partial Y}\right)^2}$$

$$\beta = \frac{1}{2}\left(\frac{\partial V}{\partial X} - \frac{\partial U}{\partial Y}\right) \tag{30}$$

6) Obtain $R$ according to the signs of $\alpha^2 - \beta^2$ and $\beta$

$$R = \begin{cases} \beta - \alpha, & \text{if } \alpha^2 - \beta^2 < 0, \beta > 0 \\ \beta + \alpha, & \text{if } \alpha^2 - \beta^2 < 0, \beta < 0 \\ 0, & \text{if } \alpha^2 - \beta^2 \geq 0 \end{cases} \tag{31}$$

7) Compute Rortex $\vec{R}$ as

$$\vec{R} = 2R\vec{r}$$

(32)

Note that if $\nabla \vec{v}$ have three real eigenvalues, it will have more than one axis of rotation $\vec{r}$. In this case, $R = 0$, so $\vec{R} = \mathbf{0}$. It will not change the determination of the vortex.

## IV. Vorticity tensor and vector decompositions

After we find the rotation axis Z, we discuss rotation in the X-Y plane.

### A. Rotation Criterion in Plane X-Y

$$\nabla V = \begin{bmatrix} \frac{\partial U}{\partial X} & \frac{\partial U}{\partial Y} \\ \frac{\partial V}{\partial X} & \frac{\partial V}{\partial Y} \end{bmatrix} = \begin{bmatrix} \frac{\partial U}{\partial X} & \frac{1}{2}\left(\frac{\partial U}{\partial Y} + \frac{\partial V}{\partial X}\right) \\ \frac{1}{2}\left(\frac{\partial U}{\partial Y} + \frac{\partial V}{\partial X}\right) & \frac{\partial V}{\partial Y} \end{bmatrix} + \begin{bmatrix} 0 & -\frac{1}{2}\left(\frac{\partial V}{\partial X} - \frac{\partial U}{\partial Y}\right) \\ \frac{1}{2}\left(\frac{\partial V}{\partial X} - \frac{\partial U}{\partial Y}\right) & 0 \end{bmatrix} = A + B \quad (33)$$

where $A$ is called symmetric deformation tensor and $B$ is anti-symmetric tensor which is usually considered as a vorticity tensor. As we discussed above, vorticity cannot be used as a quantity to describe flow rotation. The rotation criterion in a plane is given by

$$g_Z = -\frac{\partial V}{\partial X} \cdot \frac{\partial U}{\partial Y} \begin{cases} > 0 & Rotation \\ \leq 0 & Non-Rotation \end{cases} \quad (34)$$

### B. 2D Vorticity Tensor Decomposition

The 2D vorticity tensor can be further decomposed as

$$\begin{bmatrix} 0 & -\frac{1}{2}\left(\frac{\partial V}{\partial X} - \frac{\partial U}{\partial Y}\right) \\ \frac{1}{2}\left(\frac{\partial V}{\partial X} - \frac{\partial U}{\partial Y}\right) & 0 \end{bmatrix} = \begin{bmatrix} 0 & -\frac{1}{2}\left(\frac{\partial V}{\partial X} - \frac{\partial U}{\partial Y_R}\right) \\ \frac{1}{2}\left(\frac{\partial V}{\partial X} - \frac{\partial U}{\partial Y_R}\right) & 0 \end{bmatrix} + \begin{bmatrix} 0 & -\frac{1}{2}\left(-\frac{\partial U}{\partial Y_S}\right) \\ \frac{1}{2}\left(\frac{\partial U}{\partial Y_S}\right) & 0 \end{bmatrix} = C + D \quad (35)$$

where $C$ is a pure rotational tensor and $D$ is a non-rotational anti-symmetric deformation.

Here, we assume that

$\left|\frac{\partial U}{\partial Y}\right| > \left|\frac{\partial V}{\partial X}\right|$, $-\frac{\partial V}{\partial X} \cdot \frac{\partial U}{\partial Y} > 0$, let

$$\frac{\partial U}{\partial Y} = -\frac{\partial V}{\partial X} + \left(\frac{\partial V}{\partial X} + \frac{\partial U}{\partial Y}\right) = \frac{\partial U}{\partial Y_R} + \frac{\partial U}{\partial Y_S} \quad (36)$$

For example, if $\frac{\partial U}{\partial Y} = -5, \frac{\partial V}{\partial X} = 3$, let $\frac{\partial U}{\partial Y} = -\frac{\partial V}{\partial X} + \left(\frac{\partial V}{\partial X} + \frac{\partial U}{\partial Y}\right) = \frac{\partial U}{\partial Y_R} + \frac{\partial U}{\partial Y_S} = -3 - 2$, or $\frac{\partial U}{\partial Y_R} = -3, \frac{\partial U}{\partial Y_S} = -2$.

$$Rotation\ Strength = R = \frac{\partial V}{\partial X} - \frac{\partial U}{\partial Y_R} = \frac{\partial V}{\partial X} + \frac{\partial V}{\partial X} = 2\frac{\partial V}{\partial X} \quad (37)$$

If $\left|\frac{\partial U}{\partial Y}\right| < \left|\frac{\partial V}{\partial X}\right|$, $\frac{\partial V}{\partial X} \cdot \frac{\partial U}{\partial Y} < 0$, let

$$\frac{\partial V}{\partial X} = -\frac{\partial U}{\partial Y} + \left(\frac{\partial V}{\partial X} + \frac{\partial U}{\partial Y}\right) = \frac{\partial V}{\partial X_R} + \frac{\partial V}{\partial X_S} \quad (38)$$

$$R = \frac{\partial V}{\partial X_R} - \frac{\partial U}{\partial Y} = -\frac{\partial U}{\partial Y} - \frac{\partial U}{\partial Y} = -2\frac{\partial U}{\partial Y} \quad (39)$$

$$\nabla V = A + C + D \quad (40)$$

The velocity tensor is decomposed to symmetric deformation tensor $A$, anti-symmetric deformation tensor $D$ and rigid rotation tensor $C$.

## C. 3D Vorticity Tensor Decomposition

The velocity gradient tensor can be decomposed as a symmetric and a anti-symmetric tensor as usual:

$$\begin{bmatrix} \frac{\partial U}{\partial X} & \frac{\partial U}{\partial Y} & 0 \\ \frac{\partial V}{\partial X} & \frac{\partial V}{\partial Y} & 0 \\ \frac{\partial W}{\partial X} & \frac{\partial W}{\partial Y} & \frac{\partial W}{\partial Z} \end{bmatrix} = \begin{bmatrix} \frac{\partial U}{\partial X} & \frac{1}{2}\left(\frac{\partial V}{\partial X}+\frac{\partial U}{\partial Y}\right) & \frac{1}{2}\frac{\partial W}{\partial X} \\ \frac{1}{2}\left(\frac{\partial V}{\partial X}+\frac{\partial U}{\partial Y}\right) & \frac{\partial V}{\partial Y} & \frac{1}{2}\frac{\partial W}{\partial Y} \\ \frac{1}{2}\frac{\partial W}{\partial X} & \frac{1}{2}\frac{\partial W}{\partial Y} & \frac{\partial W}{\partial Z} \end{bmatrix} + \begin{bmatrix} 0 & -\frac{1}{2}\left(\frac{\partial V}{\partial X}-\frac{\partial U}{\partial Y}\right) & -\frac{1}{2}\frac{\partial W}{\partial X} \\ \frac{1}{2}\left(\frac{\partial V}{\partial X}-\frac{\partial U}{\partial Y}\right) & 0 & -\frac{1}{2}\frac{\partial W}{\partial Y} \\ \frac{1}{2}\frac{\partial W}{\partial X} & \frac{1}{2}\frac{\partial W}{\partial Y} & 0 \end{bmatrix} = A + B \quad (41)$$

**Theorem 3**. A vorticity tensor $B$ can be decomposed as a rotational anti-symmetric tensor and a non-rotational anti-symmetric tensor.

Proof: The anti-symmetric or vorticity tensor $B$ can be further decomposed as

$$B = \begin{bmatrix} 0 & -\frac{1}{2}\left(\frac{\partial V}{\partial X}-\frac{\partial U}{\partial Y}\right) & -\frac{1}{2}\frac{\partial W}{\partial X} \\ \frac{1}{2}\left(\frac{\partial V}{\partial X}-\frac{\partial U}{\partial Y}\right) & 0 & -\frac{1}{2}\frac{\partial W}{\partial Y} \\ \frac{1}{2}\frac{\partial W}{\partial X} & \frac{1}{2}\frac{\partial W}{\partial Y} & 0 \end{bmatrix} = \begin{bmatrix} 0 & -R & 0 \\ R & 0 & 0 \\ 0 & 0 & 0 \end{bmatrix} + \begin{bmatrix} 0 & -\frac{1}{2}\left(\frac{\partial V}{\partial X}-\frac{\partial U}{\partial Y}\right)+R & -\frac{1}{2}\frac{\partial W}{\partial X} \\ \frac{1}{2}\left(\frac{\partial V}{\partial X}-\frac{\partial U}{\partial Y}\right)-R & 0 & -\frac{1}{2}\frac{\partial W}{\partial Y} \\ \frac{1}{2}\frac{\partial W}{\partial X} & \frac{1}{2}\frac{\partial W}{\partial Y} & 0 \end{bmatrix} = C + D \quad (42)$$

where $C$ is a tensor corresponding to a rigid rotation with an angle speed of $R$

Tensor $D$ can be further decomposed as

$$D = \begin{bmatrix} 0 & -\frac{1}{2}\left(\frac{\partial V}{\partial X}-\frac{\partial U}{\partial Y}\right)+R & -\frac{1}{2}\frac{\partial W}{\partial X} \\ \frac{1}{2}\left(\frac{\partial V}{\partial X}-\frac{\partial U}{\partial Y}\right)-R & 0 & -\frac{1}{2}\frac{\partial W}{\partial Y} \\ \frac{1}{2}\frac{\partial W}{\partial X} & \frac{1}{2}\frac{\partial W}{\partial Y} & 0 \end{bmatrix} = \begin{bmatrix} 0 & -\frac{1}{2}\left(\frac{\partial V}{\partial X}-\frac{\partial U}{\partial Y}\right)+R & 0 \\ \frac{1}{2}\left(\frac{\partial V}{\partial X}-\frac{\partial U}{\partial Y}\right)-R & 0 & 0 \\ 0 & 0 & 0 \end{bmatrix} +$$

$$\begin{bmatrix} 0 & 0 & -\frac{1}{2}\frac{\partial W}{\partial X} \\ 0 & 0 & 0 \\ \frac{1}{2}\frac{\partial W}{\partial X} & 0 & 0 \end{bmatrix} + \begin{bmatrix} 0 & 0 & 0 \\ 0 & 0 & -\frac{1}{2}\frac{\partial W}{\partial Y} \\ 0 & \frac{1}{2}\frac{\partial W}{\partial Y} & 0 \end{bmatrix} = E + F + G \quad (43)$$

According to our rotation criteria, $F$ and $G$ have no rotation since $g_X = \frac{\partial W}{\partial Y} \cdot \frac{\partial V}{\partial Z} = \frac{\partial W}{\partial Y} \cdot 0 = 0$ and $g_Y = \frac{\partial W}{\partial X} \cdot \frac{\partial U}{\partial Z} = \frac{\partial W}{\partial X} \cdot 0 = 0$.

About tensor $E$, we have

$$-\frac{1}{2}\left(\frac{\partial V}{\partial X}-\frac{\partial U}{\partial Y}\right) + R = -\frac{1}{2}\left(\frac{\partial V}{\partial X}-\frac{\partial U}{\partial Y_R}\right) + R + \frac{\partial U}{\partial Y_S} = 0 + \frac{\partial U}{\partial Y_S} = \frac{\partial U}{\partial Y_S},$$

$$E = \begin{bmatrix} 0 & -\frac{1}{2}\left(\frac{\partial V}{\partial X} - \frac{\partial U}{\partial Y}\right) + R & 0 \\ \frac{1}{2}\left(\frac{\partial V}{\partial X} - \frac{\partial U}{\partial Y}\right) - R & 0 & 0 \\ 0 & 0 & 0 \end{bmatrix} = \begin{bmatrix} 0 & \frac{\partial U}{\partial Y_S} & 0 \\ -\frac{\partial U}{\partial Y_S} & 0 & 0 \\ 0 & 0 & 0 \end{bmatrix},$$

$$g_Z = \frac{\partial U}{\partial Y_S} \cdot \frac{\partial V}{\partial X} = \frac{\partial U}{\partial Y_S} \cdot 0 = 0 \tag{44}$$

Therefore, we prove that tensor $D$ has no rotation. The anti-symmetric tensor $B$ can be decomposed to a rotational anti-symmetric tensor $C$ and a non-rotational anti-symmetric tensor $D$.

### D. Uniqueness of the Tensor Decomposition of $\nabla V = A + C + D$

**Theorem 4.** The tensor decomposition of $B = C + D$ is unique.

Proof. If we add any number $\varepsilon$ to $R$, we will obtain another decomposition of tensor $B$

$$B = \begin{bmatrix} 0 & -R - \varepsilon & 0 \\ R + \varepsilon & 0 & 0 \\ 0 & 0 & 0 \end{bmatrix} + \begin{bmatrix} 0 & -\frac{1}{2}\left(\frac{\partial V}{\partial X} - \frac{\partial U}{\partial Y}\right) + R + \varepsilon & -\frac{1}{2}\frac{\partial W}{\partial X} \\ \frac{1}{2}\left(\frac{\partial V}{\partial X} - \frac{\partial U}{\partial Y}\right) - R - \varepsilon & 0 & -\frac{1}{2}\frac{\partial W}{\partial Y} \\ \frac{1}{2}\frac{\partial W}{\partial X} & \frac{1}{2}\frac{\partial W}{\partial Y} & 0 \end{bmatrix} = C_2 + D_2 \tag{45}$$

Where $D_2$ is non-rotational

$$D_2 = \begin{bmatrix} 0 & -\frac{1}{2}\left(\frac{\partial V}{\partial X} - \frac{\partial U}{\partial Y}\right) + R + \varepsilon & -\frac{1}{2}\frac{\partial W}{\partial X} \\ \frac{1}{2}\left(\frac{\partial V}{\partial X} - \frac{\partial U}{\partial Y}\right) - R - \varepsilon & 0 & -\frac{1}{2}\frac{\partial W}{\partial Y} \\ \frac{1}{2}\frac{\partial W}{\partial X} & \frac{1}{2}\frac{\partial W}{\partial Y} & 0 \end{bmatrix} = \begin{bmatrix} 0 & \frac{\partial U}{\partial Y_S} + \varepsilon & 0 \\ -\frac{\partial U}{\partial Y_S} - \varepsilon & 0 & 0 \\ 0 & 0 & 0 \end{bmatrix} + \begin{bmatrix} 0 & 0 & -\frac{1}{2}\frac{\partial W}{\partial X} \\ -\frac{\partial U}{\partial Y_S} - \varepsilon & 0 & 0 \\ \frac{1}{2}\frac{\partial W}{\partial X} & 0 & 0 \end{bmatrix} +$$

$$\begin{bmatrix} 0 & 0 & 0 \\ 0 & 0 & -\frac{1}{2}\frac{\partial W}{\partial Y} \\ 0 & \frac{1}{2}\frac{\partial W}{\partial Y} & 0 \end{bmatrix} = E_2 + F_2 + G_2 \tag{46}$$

If $\varepsilon = \gamma_1 \frac{\partial U}{\partial Y}$, there will be no change to $R$ and $C$, but if $\varepsilon = \gamma_2 \frac{\partial V}{\partial X}$,

$$= \begin{bmatrix} 0 & \frac{\partial U}{\partial Y_S} + \gamma_2 \frac{\partial V}{\partial X} & 0 \\ -\left(\frac{\partial U}{\partial Y_S} + \gamma_2 \frac{\partial V}{\partial X}\right) & 0 & 0 \\ 0 & 0 & 0 \end{bmatrix} \tag{47}$$

$$g_Z = -\frac{\partial U}{\partial Y_S} \cdot \varepsilon \neq 0$$

$E_2$ and then $D_2$ become rotational, which is a contradiction to our assumption that $D_2$ is non-rotational. Therefore, we must have $\varepsilon \equiv 0$.

Theorem 4 clearly shows that the vorticity tensor decomposition $\boldsymbol{B} = \boldsymbol{C} + \boldsymbol{D}$ is unique. $\vec{R}$ obtained from $\boldsymbol{C}$, which is now called Rortex, is the unique mathematical definition of flow rotation or vortex vector. All other vortex definitions or vortex identifications must be inaccurate or inappropriate if they are not identical to $\vec{R}$.

## V. Vorticity Vector Decomposition

The tensor decomposition of anti-symmetric tensor or vorticity tensor will lead to an extremely important vector decomposition which is the vorticity vector decomposition.

Theorem 5. Vorticity vector can be decomposed to a rotational Rortex vector and non-rotational shear vector or $\nabla \times \boldsymbol{V} = \vec{R} + \vec{S}$ (Fig. 1).

Proof: Assume $\vec{dl}$ is an arbitrarily selected real vector,

$$2\boldsymbol{B} \cdot \vec{dl} = -\vec{dl} \times (\nabla \times \boldsymbol{V}) = 2\boldsymbol{C} \cdot \vec{dl} + 2\boldsymbol{D} \cdot \vec{dl} = -\vec{dl} \times \vec{R} - \vec{dl} \times \vec{S}$$

$$\vec{dl} \times (\nabla \times \boldsymbol{V}) = \vec{dl} \times \vec{R} + \vec{dl} \times \vec{S} = \vec{dl} \times (\vec{R} + \vec{S}) \qquad (48)$$

Since $\vec{dl}$ is arbitrarily selected, we will have

$$\nabla \times \vec{V} = \vec{R} + \vec{S} \qquad (49)$$

This proof clearly shows vorticity vector $\nabla \times \vec{V}$ cannot represent flow rotation and only Rortex $\vec{R}$ can represent the flow rotation which plays a critical role in turbulent flow.

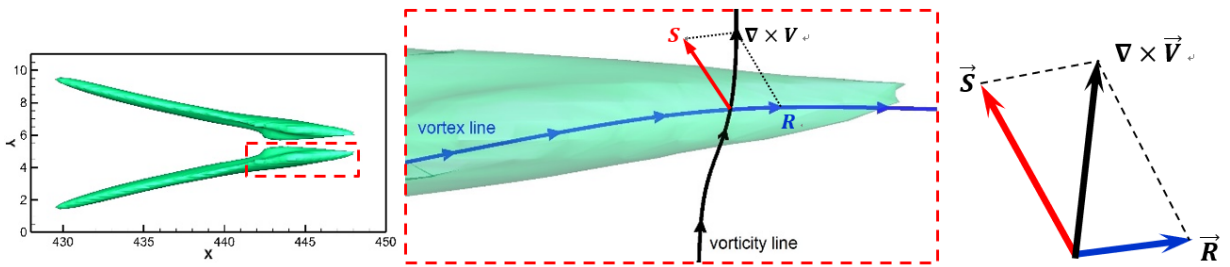

FIG. 1. Vorticity vector is decomposed to Rortex vector and anti-symmetric shear vector.

## VI. COMPUTATIONAL RESULTS

### A. 2D Couette flow and 2D rigid rotation

For 2D flow (Fig. 2), there is only one rotational axis which is the Z-axis. The criterion of fluid rotation is given by $g_Z = -\frac{\partial U}{\partial Y} \cdot \frac{\partial V}{\partial X} > 0$. For the Couette flow, $U = \omega Y, V = 0$ and for 2D rigid rotation, $U = \omega Y, V = -\omega X$. Both cases have vorticity $\omega_Z = \frac{1}{2}\left(\frac{\partial V}{\partial X} - \frac{\partial U}{\partial Y}\right)$, but the Couette flow has $g_Z = -\frac{\partial U}{\partial Y} \cdot \frac{\partial V}{\partial X} = -\frac{\partial U}{\partial Y} \cdot 0 = 0$ with no rotation, but the 2D rigid rotation has $g_Z = -\frac{\partial U}{\partial Y} \cdot \frac{\partial V}{\partial X} = \omega \cdot \omega = \omega^2 > 0$ with rotation.

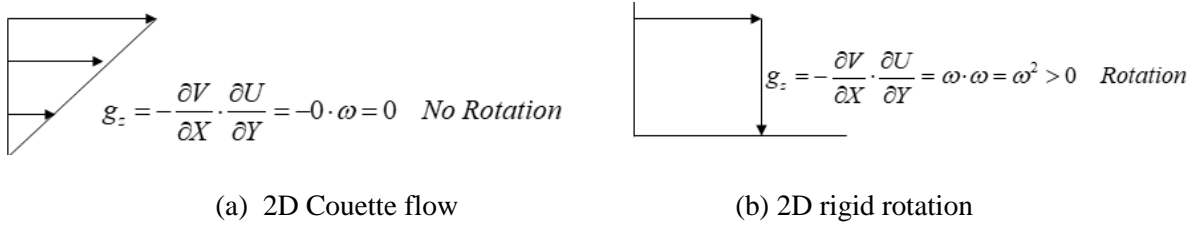

(a) 2D Couette flow  (b) 2D rigid rotation

FIG. 2. 2D Couette flow and 2D rigid rotation.

## B. 3D boundary layer transition on a flat plate

We use the new defined to identify the vortex structure in late boundary layer transition on a flat plate (Fig. 3) which is simulated by a DNS code called DNSUTA.[25] The DNS was conducted with near 60 million grid points and over 400,000 time steps at a free stream Mach number of 0.5. Being different from other vortex identification methods which only use iso-surface to detect the vortex structure, this new defined Rortex can use not only Rortex iso-surface but also Rortex vector field and Rertex lines to identify the vortex structure. Figs 4 and 5 shows both the Rortex iso-surface and the Rortex vector field for the first vortex ring and later vortex rings respectively, which clearly show the Rortex vector filed coincide with the Rortex iso-surface while the vorticity lines could be orthogonal to vortex surface or penetrate the vortex surface. Figs 6 and 7 depict the Rortex lines for the hairpin vortices and ring-like vortices respectively, which clearly show that the Rortex lines coincide with the Rortex tubes and there should be no Rortex lines leakage from the Rortex tubes. However, the Rortex tube could be generated, developed, and ended inside the flow field, which is very different from vorticity tubes. These examples clearly show that vorticity cannot be used to identify the vortex structure properly, but the newly defined Rortex can do properly by using not only Rortex iso-surface but also Rortex vectors, Rortex lines, and Rortex tubes. The reason is quite simple that vorticity is not vortex vector as $\nabla \times \vec{V} = \vec{R} + \vec{S}$. Vorticity can be used to show the vortex structure properly only when $\vec{S} = 0$ where the fluid become rigid.

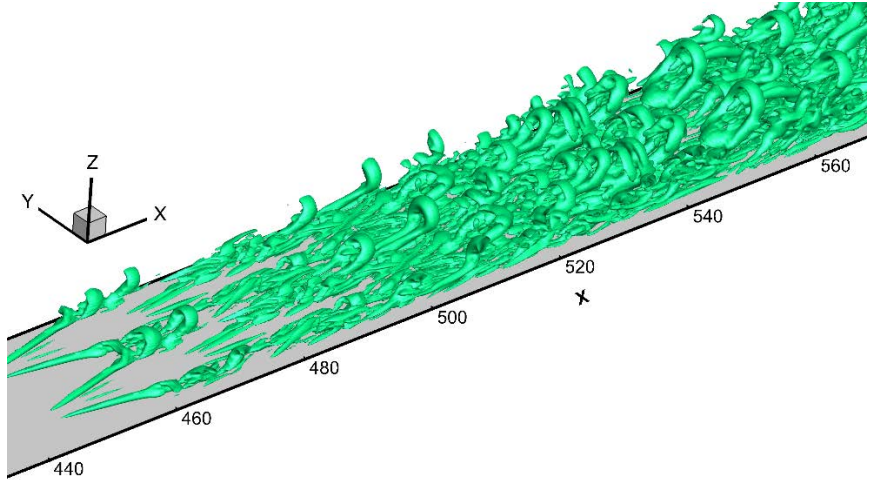

FIG. 3. Vortex structure of late boundary layer transition with iso-surface of |R|=0.75.

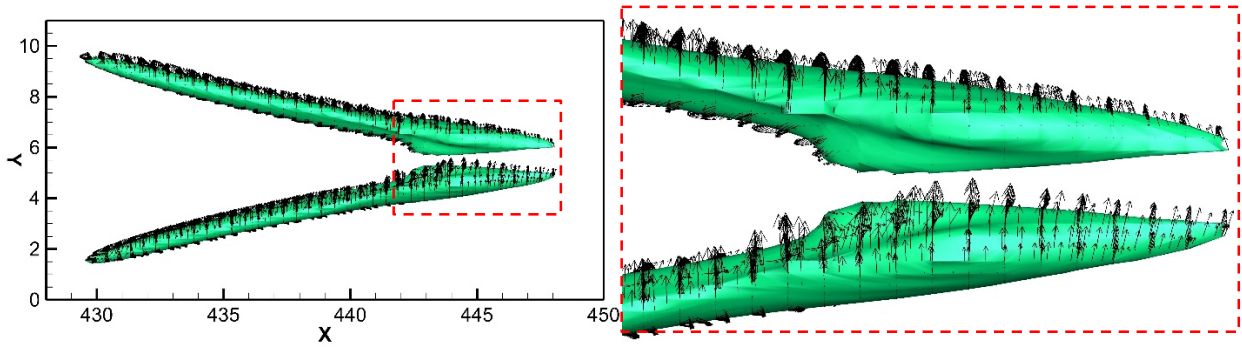

(a) Vorticity vector field and vortex surface.

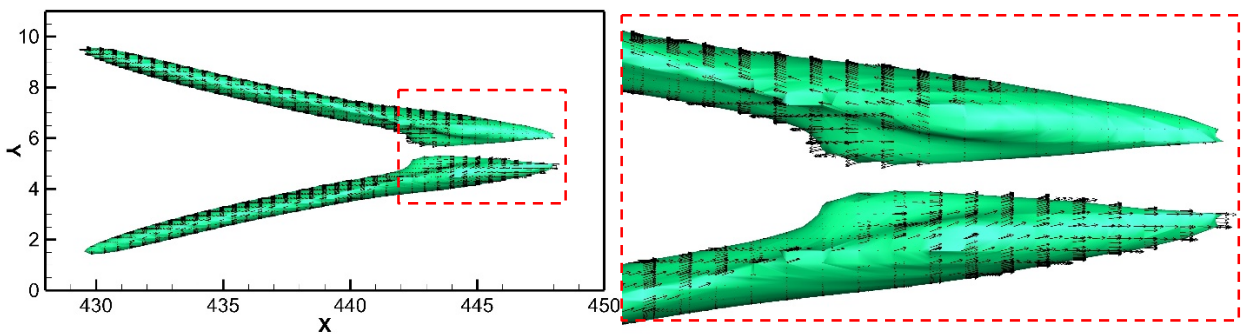

(b) Rortex vector field and vortex surface.

FIG. 4. Comparison of Vorticity and Rortex.

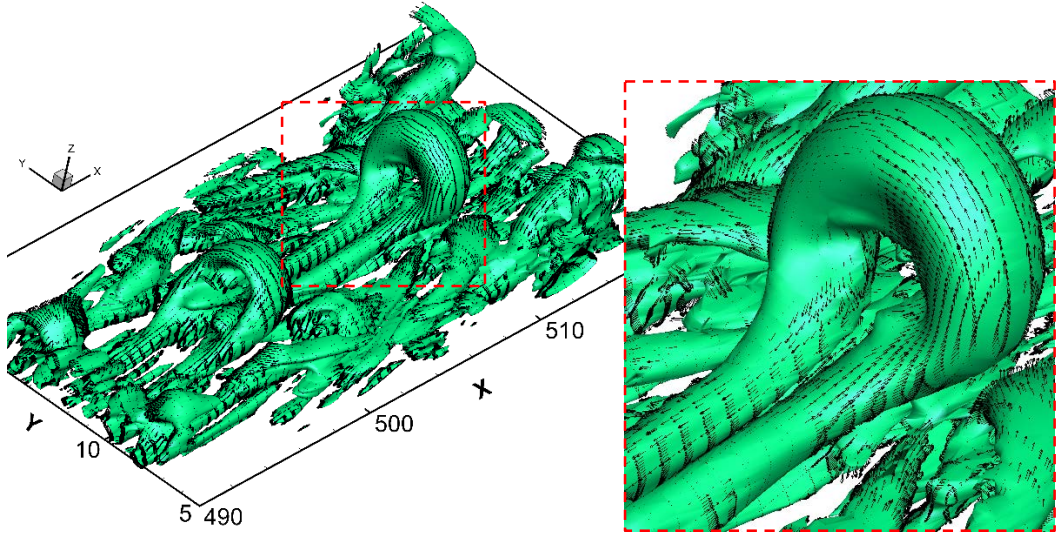

FIG. 5. Rortex vector field on the R iso-surface of the first vortex ring.

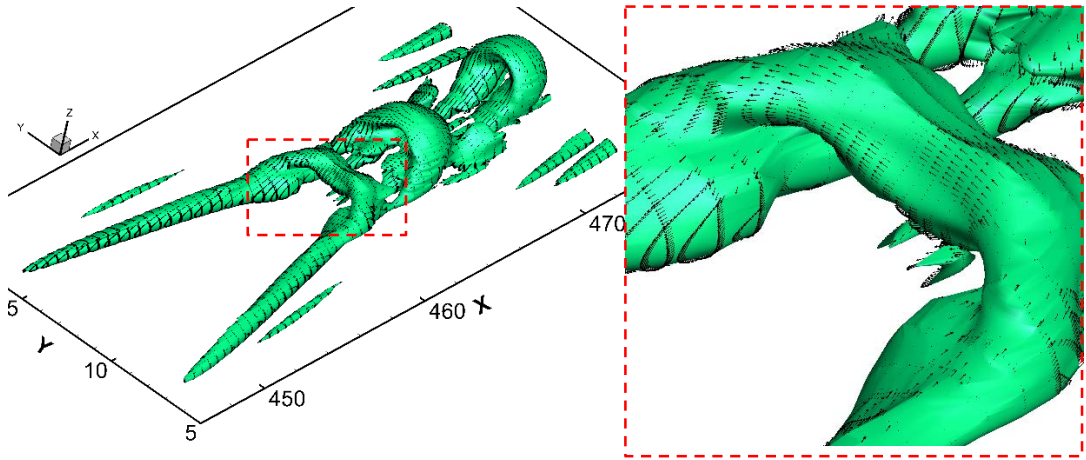

FIG. 6. Rortex vector field on the R iso-surface of the later vortex rings.

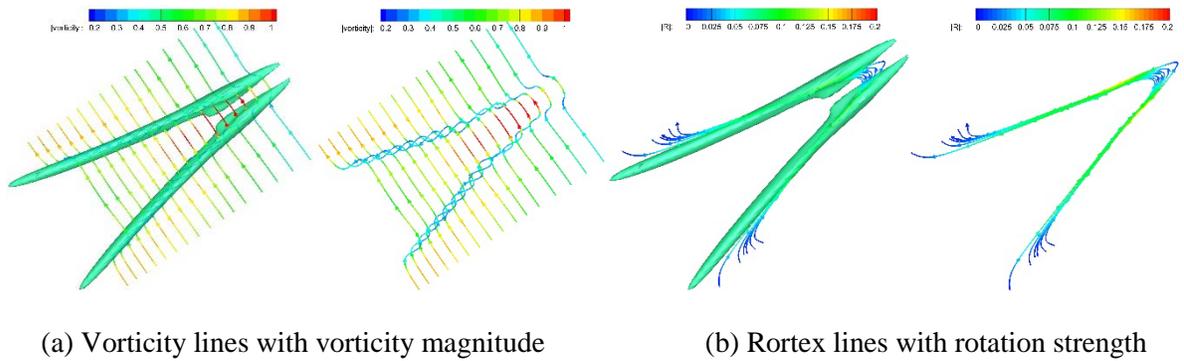

(a) Vorticity lines with vorticity magnitude     (b) Rortex lines with rotation strength

FIG. 7. Comparison of vorticity lines and Rortex lines for the Lambda vortex.

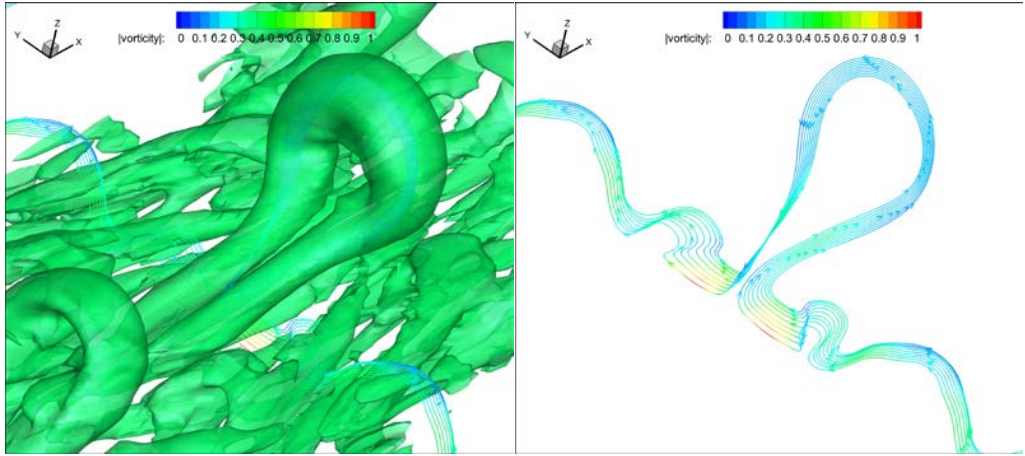

(a) Vorticity lines with vorticity magnitude.

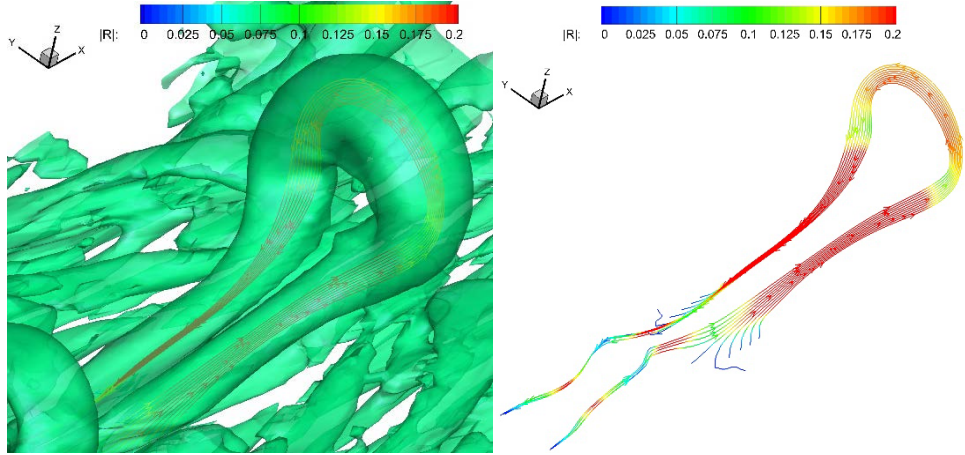

(b) Rortex lines with rotation strength.

FIG. 8. Comparison of vorticity lines and Rortex lines for the first ring vortex.

## VII. CONCLUSIONS

1. An accurate mathematical definition of vortex vector was given by our previous work. Since this new definition is directly related to fluid rigid rotation, we give a new name "Rortex" with more mathematical analysis given in this paper.

2. The existence of Rortex is mathematically proved.

3. Not only the iso-surface of Rortex can be used, but also the vortex vector can also be used to show the vortex structure including Rortex vector field, Rortex Lines, Rortex tubes, and Rortex surfaces.

4. Rortex lines are parallel to vortex structure unlike vorticity lines which can penetrate vortex structure.

5. Not only vortex strength iso-surface which has same vortex strength can be applied to describe the vortex structure, but also different vortex strength along the Rortex lines can be applied to do analysis on physics of vortex and/or turbulence generation.

6. It is very convenient and fast to calculate Rortex vector including the direction and strength （we need about one minute in a regular laptop to do the post processing for one-time step with 60 million of grids).

7. The anti-symmetric tensor of the velocity gradient or vorticity tensor can be decomposed to a rotational part and anti-symmetric deformation part.

8. Vorticity vector can be decomposed to a Rortex part and shear part, i.e. $\nabla \times \vec{V} = \vec{R} + \vec{S}$, which represent rotational part and non-rotational part respectively.

9. The newly defined Rortex is a physical quantity closely related to turbulence structure and strength. This will provide a new powerful tool to study vortex dynamics and physics of turbulence generation and development.

**ACKNOWLEDGEMENTS**

This work was supported by the Department of Mathematics at University of Texas at Arlington and AFOSR grant MURI FA9559-16-1-0364. The authors are grateful to Texas Advanced Computing Center (TACC) for providing computation hours. This work is accomplished by using Code DNSUTA which was released by Dr. Chaoqun Liu at University of Texas at Arlington in 2009.